\documentclass[twocolumn]{revtex4-1}

\usepackage{graphicx}
\usepackage{epsfig}

\usepackage[english]{babel}

\usepackage{amsmath,amssymb}
\usepackage{textcomp}

\usepackage{caption}
\usepackage{xcolor}
\usepackage[colorlinks=true]{hyperref}
\hypersetup{linkcolor=[rgb]{0.6 0 0}, citecolor=[rgb]{0 0.4 0}, urlcolor=[rgb]{0 0 0.6}}

\begin{document}

\title{Spiral structures in helical magnets}

\author{A.B. Borisov}
\author{F.N. Rybakov}
\email{f.n.rybakov@gmail.com}

\affiliation{Institute of Metals Physics, Urals Branch of the Russian Academy of Sciences, Ekaterinburg, 620990,
Russia}

\date{September 17, 2012}

\begin{abstract}
Structure and properties of two-dimensional spiral textures in helical ferromagnets have been studied. In these magnetic mediums have been predicted new types of periodical structures - spiral lattices.
\end{abstract}

\maketitle

Different spin structures in magnetic systems are the matter of intensive theoretical and experimental researches. Last years a study of spin textures in non-centrosymmetric helicoidal ferromagnetic crystals have aroused doubtless interest. Here, owing to Dzyaloshinskii-Moriya interaction, the ground state is one-dimensionally modulated phase - helicoid \cite{bib:DzHelicPart3}. Thereby such spin ordering with edge magnetic dislocations was discovered in cubic crystal Fe$_{0.5}$Co$_{0.5}$Si by methods of transmission electron microscopy in the paper \cite{bib:Uchida2006}. In the following paper \cite{bib:Uchida2008} the same methods were used for observing spiral (Swiss-roll-like vortex) in helimagnet FeGe with continual heating of the sample from 85K to 200K. Finally, skyrmion lattice, which was predicted many years ago \cite{bib:BogdanovVortex1,bib:BogdanovVortex3}, was found recently \cite{bib:Yu1}. To the present date there is no theoretical and numerical analysis of two-dimensional structures in helimagnets except skyrmions and their lattices.
 
In this paper by using analytical and numerical methods for two-dimensional model we research the structure and properties of spiral textures, which were recently discovered experimentally \cite{bib:Uchida2008}. An analytic formula for the structure of spirals outside the corn was calculated. It matches with numerical calculations and it gave an opportunity to research their properties including behaviour in the magnetic field. Existence of single structures in condensed matters (dislocations, disclinations, vortexes, skyrmions, etc) is as the rule accompanied by the formation of their periodical structures. We are predicting new types of periodical structures in helical ferromagnets - the Archimedean spiral lattices.  

\begin{figure}[!htb]
\includegraphics[width=1.0\columnwidth]{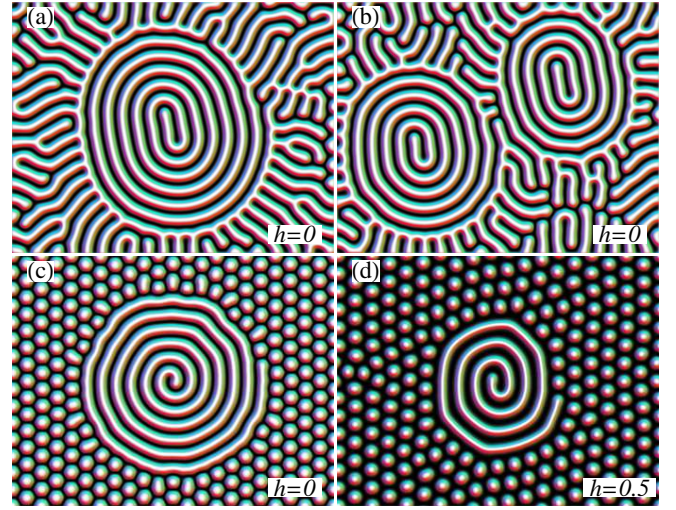}
\caption{Calculated Archimedean spirals surrounded by a different structures. (a),(b) AS surrounded by a labyrinth structure. (c) AS surrounded by a skyrmion lattice in zero field; (d) - in the filed $h=0.5$.}
\label{F:fig1}
\end{figure}

The energy of the helical ferromagnetic cubic crystals (such as FeGe, MnSi \cite{bib:BakJensen}) includes contributions from the exchange interaction, external magnetic field $H_z$ and Dzyaloshinskii-Moriya interaction:
\begin{multline}
E_M=\frac{\alpha M^2}{2} \int ({\partial _{i}}{\bf n})^2d{\bf r} + H_{z}\,M \int(1 - n_z)d{\bf r} + {} \\
{D M^2} \int {\bf n}\cdot\left(  {\bf \nabla}\times{\bf n}\right)d{\bf r}  ,\label{eq:Energy}
\end{multline}
where ${\bf n}=(n_x,n_y,n_z)$ is a unit vector of magnetization:
\begin{equation}
{\bf n} = (sin \theta cos \phi, sin \theta sin \phi, cos \theta ), \quad {\bf n}^2 = 1  ,\label{eq:nparam}
\end{equation}
$M$ - saturation magnetization, $\alpha$ - exchange constant, $D$ - Dzyaloshinskii constant.    

The Euler-Lagrange equations for energy functional (\ref{eq:Energy}) in polar coordinate system $(r,\varphi)$:
\begin{multline}
\frac{4 \pi (sin\theta)^2}{L_D}\left( {\phi_{'r}} cos(\varphi - \phi) - \frac{\phi_{'\varphi}}{r} sin(\varphi - \phi)   \right) + {}   \\ 
{}+\Delta\theta - \frac{sin(2\theta)}{2} \left( {\phi_{'r}}^2 + \frac{{\phi_{'\varphi}}^2}{r^2}  \right) - \frac{4 \pi^2 h}{{L_D}^2}sin\theta = 0			, \label{eq:EL1}
\end{multline}
\begin{multline} 
\frac{4 \pi sin\theta}{L_D}\left(\frac{{\theta_{'\varphi}}}{r}sin(\varphi - \phi) -  {\theta_{'r}}cos(\varphi - \phi)   \right)  +{} \\
 {}+ sin\theta\Delta\phi + 2 cos\theta \left( \frac{{\theta_{'\varphi}}{\phi_{'\varphi}}}{r^2} +  {\theta_{'r}}{\phi_{'r}}  \right) = 0		,\label{eq:EL2}
\end{multline}
where $\Delta$ is the Laplace operator and parameters
\begin{equation}
h = H_z / H_D, \quad H_D = D^2 M / \alpha, \quad L_D = 2 \pi \alpha / D
\end{equation}
are the same as in the paper \cite{bib:ChiralSkyrmionicMatter}.

The method of energy functional (\ref{eq:Energy}) minimization was used for solving equations (\ref{eq:EL1})-(\ref{eq:EL2}).  
Calculations were made on the high density grid: 2048$\times$1024. So that the angle between the two directions of a vector in the adjacent nodes did not exceed $10^{\circ}$. 
The nonlinear conjugate gradient method with additive penalty functions, which are required for vector norm fixation (\ref{eq:nparam}), was used. 
To control the result accuracy the first and the second derivatives were calculated and then directly discrepancy in the equations (\ref{eq:EL1})-(\ref{eq:EL2}). 
Algorithm was realized on the Nvidia CUDA architecture using video cards microprocessors (GPU) for massive parallel calculations and visualization in real time. The usage of such technologies allow to increase sufficiently the speed/accuracy of a number of numerical problems, including micro-magnetic ones \cite{bib:MuMax}. 
As a result different types of spiral structures were found. They will be described further. Moreover, the analysis of obtained numerical results gave us an opportunity to make analytical research of their structure and properties too.  

We found asymptotic spiral structure of solutions of (\ref{eq:EL1})-(\ref{eq:EL2}). Thus for $h=0$:
\begin{equation}
\begin{cases}
\phi \rightarrow  \frac{\pi}{2} + \varphi  - \frac{N L_D}{2\pi r} + \mathcal{O}(\frac{1}{r^2}),&\\
\theta \rightarrow  c_1 + c_2 log(r) - \frac{2\pi}{L_D}r +{} \\
{} \quad + N\,\varphi + \mathcal{O}(\frac{1}{r}), &
\end{cases} \quad (r\rightarrow\infty)
\end{equation}
where $N\in\mathbb{Z}$. The origin of logarithmic term is tied to the exchange spirals \cite{bib:BorisovExSp}. In the case $D \neq 0$ it is necessary for minimization (\ref{eq:Energy}) to set arbitrary constant $c_2=0$. Then solutions for $n_z=cos(\theta)$ are 2N-spiral domains, which are divided by Archimedean spirals. It is possible to show, that when $h>0$ asymptotic solution is:
\begin{equation}
\begin{cases}
\phi \rightarrow  \frac{\pi}{2} + \varphi - \frac{N L}{2\pi r} + \mathcal{O}(\frac{1}{r^2}),&\\
\theta \rightarrow \pi - 2\,am \Big(F(\frac{\pi - c1}{2},\varkappa) +{} \\
{} \quad + \frac{2 K}{L}r - \frac{K\,N}{\pi}\varphi +  \mathcal{O}(\frac{1}{r}),\, \varkappa \Big), &
\end{cases} (r\rightarrow\infty)  \label{eq:asym}
\end{equation}
where $am$ - Jacobi amplitude, $F$ - elliptic integral of the first kind,  
$K=K(\varkappa)$ - complete elliptic integral of the first kind. Elliptic modulus $\varkappa$ depends only on $h$ and determines by equation
\begin{equation}
\pi \varkappa - 4\sqrt{h}E=0,
\end{equation}
where $E=E(\varkappa)$ complete elliptic integral of the second kind. 
It is noteworthy that the period of spiral turns 
\begin{equation}
L=\frac{L_D}{\pi \sqrt{h}} \, \varkappa \, K=L_D \frac{4 E K}{\pi^2}
\end{equation}
and coincides with the period of helicoid \cite{bib:DzHelicPart3,bib:Izumov1,bib:Izumov2}.

If $N=0$, then solution conforms to the extreme case of \emph{$k\pi$-vortices} \cite{bib:BogdanovHubert99}. For $|N|=1$, level lines $n_z=const$ are the Archimedean spiral (AS) (if $|N|>1$, then spiral is multiarm). 

By integrating the functional (\ref{eq:Energy}) for the leading terms (\ref{eq:asym}), we find the energy, which is contained in $k$ turns ($k>1$):
\begin{align}
& E_{AS} \simeq E_0\cdot(\varepsilon_{c} + g_1 k^2 + g_2 log(k)),\label{eq:EA}\\
& g_1 = 8 \pi  K^2 \left(\varkappa ^2-1\right), \\
& g_2 = \frac{4 \pi  E \left(2-\varkappa ^2\right)}{3 K \varkappa ^4}+\frac{8 \pi  \left(\varkappa ^2-1\right)}{3 \varkappa ^4},\label{eq:g2}
\end{align}
where $E_0=\alpha M^2$, $\varepsilon_{c}$ - dimensionless quantity of the core energy, which depends only on $h$ and $N$. Core structure can not be determined by asymptotic method and energy $\varepsilon_{c}$ was obtained by numerical method. 

It follows from (\ref{eq:g2}) that the direction of AS curling and a number of arms does not influence spiral energy density at a distance from the core. As a result of this we studied only the simplest case $N=1$. Such kind of AS was observed in the experiment \cite{bib:Uchida2008}.

As we can see from the numerical calculation, spiral textures as metastable states can exist surrounded by labyrinth structures (fig.\ref{F:fig1}a,b) or surrounded by skyrmions (fig.\ref{F:fig1}c,d), like static spiral domains (SD), surrounded by labyrinth domain structure \cite{bib:Ges1990} or by bubble lattice \cite{bib:Mamalui1996,bib:Mamalui2008} in garnet ferrite films. In these films SDs may contain up to 30 turns \cite{bib:LogRand} and being observed only as dynamical structures in some samples \cite{bib:Kandaurova}. Magnetic dipole interactions play the key role in forming static SDs \cite{bib:BorJal}, but for the Archimedean spiral in chiral magnets the same as for other solitons with typical sizes about $L_D$, taking into account the influence of these long-ranging forces is not so critical \cite{bib:Kiselev}.   

On the fig.\ref{F:fig1} we can see how does the AS changes its form a little under the influence of the surrounding. Similar spirals for the director field can be observed in the liquid crystals films \cite{bib:Gillia,bib:Mitov,bib:Ribiere}. However, director field is highly inhomogeneous over the film thickness and contains topological defects \cite{bib:Oswald}. AS vector field is continuous everywhere and does not contain any singularities but it is topologically nontrivial. Localized AS (fig.\ref{F:fig1}c,d) can be smoothly deformed into Belavin-Polyakov vortex \cite{bib:BP} with unit topological charge.  For instance, if the field will be increased, AS on the fig.\ref{F:fig1}d will lose stability and become the same skyrmion as the others around.    

\begin{figure}[!htb]
\includegraphics[width=1.0\columnwidth]{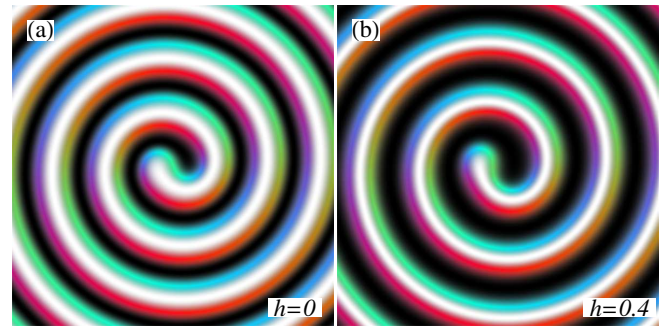}
\caption{Calculated structure of pure AS. (a) - in zero field; (b) - in the field $h=0.4$.}
\label{F:fig2}
\end{figure}

\begin{figure}[!htb]
\includegraphics[width=1.0\columnwidth]{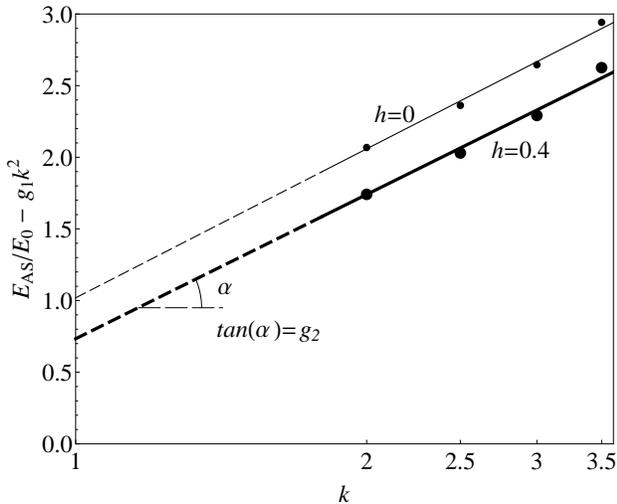}
\caption{AS energy dependence (relative to the helicoid) from the number of turns, in logarithmic scale. Thin line - in zero field, thick line - in the field $h=0.4$.}
\label{F:fig3}
\end{figure}

For studying the structure of undistorted AS core we used the formula (\ref{eq:asym}) for fixing boundary conditions. Calculated AS structure is presented on the fig.\ref{F:fig2}. 
It is noticeable that when the external field increases, then period increases too and areas, which are magnetized against the field become thinner. Dependence on the fig.\ref{F:fig3} shows that (\ref{eq:EA}) is suitable even for the case of only two turns. 
Values for $k=3.5$ does not match a little to the dependency, but it was made for being closer to the limit of accuracy of the computational approach. Required accuracy grows in proportion to $k^2$ and for $k>3$ is already comparable with calculating accuracy. By using a straight line approximation (fig.\ref{F:fig3}), according to (\ref{eq:EA}), determining appropriate value for $k=1$, it is possible to calculate the core energy. The obtained dependence
\begin{equation}
\varepsilon_{c} \simeq 1.0 - 0.6\,h  \label{eq:Ecore}
\end{equation}
in common with the formula (\ref{eq:EA}) may be used for high accurate calculation of AS energy with any quantity of turns.

\begin{figure}[!htb]
\includegraphics[width=1.0\columnwidth]{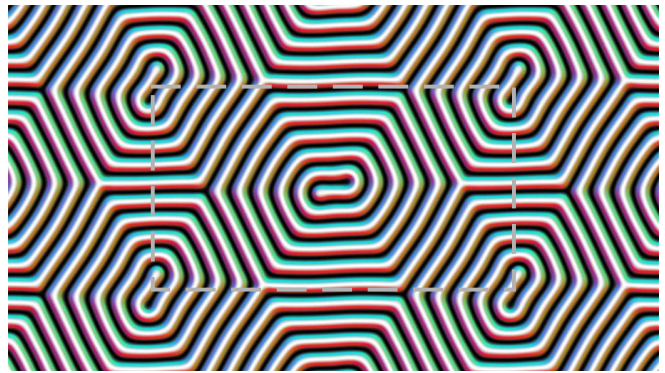}
\caption{Calculated ASL structure with five turns in zero field. Selected domain is the computational grid area, which is the unit cell.}
\label{F:fig4}
\end{figure}

New equilibrium state - hexagonal lattice of Archimedean spirals (ASL) with different number of turns was found using the method of average energy density minimization with periodical boundary conditions. Such structures remain stable in small perturbations and do not transform into helicoid. Superposition of identical precomputed solitary AS was chosen for the initial state in such a way that four centres were set in the corners of rectangular cell (fig.\ref{F:fig4}) and one - in the middle.
In the process of calculation turns changed shape to hexagonal and the cores of spirals distorted and slightly elongated. The central core rotated noticeably relative to the corner ones.
On the fig.\ref{F:fig4} a structure of equilibrium lattice with five turns is shown. 
Though this phase is similar to SD lattices, which were observed in a certain kind of ferrimagnetic films \cite{bib:Mamalui2001,bib:Mamalui2008}, its existence is due to the local Dzyaloshinskii-Moriya interaction but not to the long-ranging magnetic dipole forces.

We made a numerical analysis of the energy of different two-dimensional structure with zero temperature and it showed the following. In zero magnetic field the average density of energy (in units $E_0/L_D^2$) of Dzyaloshinskii helicoidal phase $\sigma_{D}=-19.74$, for ASL - $\sigma_{AS}=-19.58$, for skyrmion lattice -  $\sigma_{S}=-18.31$. Negative sign in expression is tied with negative sign of energy of Dzyaloshinskii-Moriya interaction. 
In the magnetic field when $0\leqslant h<0.2$ values of these energies are ordered as $\sigma_{D}<\sigma_{AS}<\sigma_{S}$, ASL loses its stability when $h\approx 0.2$. 
That is why ASL may be a basic state only when temperatures are different from zero, in the range of fields $0\leqslant h<0.2$. 
As the number of ASL degrees of freedom is much higher than the number of one-dimensional helicoid degrees of freedom then its free energy may be smaller. For the verification of this statement we should calculate system free energy (with intermediary temperatures where AS are actually observed \cite{bib:Uchida2008}). At present, these calculations cannot be fully completed neither theoretically nor numerically. ASL as a metastable state may be obtained in periodical magnetic field as it is used for creation of SD lattices \cite{bib:Mamalui2001}.                

The authors thank  A.N. Bogdanov, U.K. R\"o\ss ler, M.V. Sadovskii and N.S. Kiselev  for useful discussions.

\section*{Appendix}
\begin{figure}[!htb]
\includegraphics[width=1.0\columnwidth]{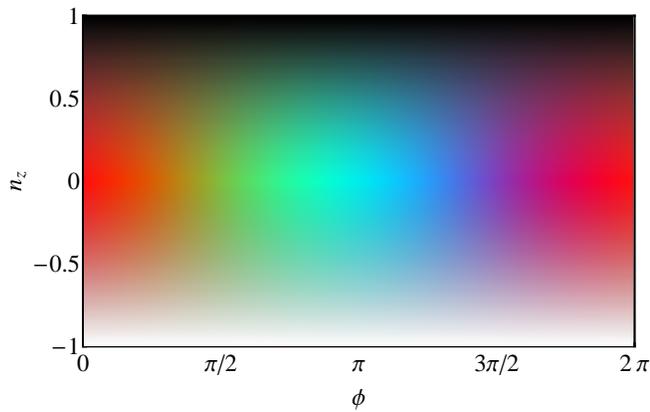}
\caption{Color scheme, which determines the direction of the magnetization vector in three dimensions.}
\label{F:fig5}
\end{figure}

\end{document}